\documentclass[twocolumn,showpacs,preprintnumbers,amsmath,amssymb]{revtex4}
\usepackage{graphicx}
\usepackage{color}

\def\rez{$^{\rm a}$}
\def\dub{$^{\rm b}$}
\def\gsi{$^{\rm c}$}
\def\fra{$^{\rm d}$}
\def\mpi{$^{\rm e}$}
\def\hei{$^{\rm f}$}
\def\wei{$^{\rm g}$}
\def\mue{$^{\rm h}$}
\def\sun{$^{\rm i}$}
\def\bnl{$^{\rm j}$}
\def\cer{$^{\rm k}$}

\def\pt{$p_t$}
\def\K0{K$^0$}

\def\agev{~A~GeV\!/c}
\def\sg{$\sigma_{\rm G}$}
\def\degrees{$^{\rm o}$}

\def\rout{R_{\rm out}}
\def\rside{R_{\rm side}}
\def\rlong{R_{\rm long}}
\def\routside{R_{\rm os}}
\def\routlong{R_{\rm ol}}
\def\rsidelong{R_{\rm sl}}

\def\Rout{$\rout$}
\def\Rside{$\rside$}
\def\Rlong{$\rlong$}

\def\Routlong{$\routlong$}
\def\Rsidelong{$\rsidelong$}

\def\qside{$q_{\rm side}$}

\def\pip{$\pi^+$}
\def\pim{$\pi^-$}

\begin{document}

\title{Azimuthal dependence of pion source radii \\ 
in Pb+Au collisions at 158\agev}

\author{
D.~Adamov\'{a}\rez,
G.~Agakichiev\dub,
A.~Andronic\gsi,
D.~Anto\'{n}czyk\fra,
H.~Appelsh\"{a}user\fra,
V.~Belaga\dub,
J.~Biel\v{c}\'{i}kov\'{a}\mpi$^,$\hei,
P.~Braun-Munzinger\gsi,
O.~Busch\hei,
A.~Cherlin\wei,
S.~Damjanovi\'{c}\hei,
T.~Dietel\mue,
L.~Dietrich\hei,
A.~Drees\sun,
W.~Dubitzky\hei,
S.\thinspace I.~Esumi\hei,
K.~Filimonov\hei,
K.~Fomenko\dub,
Z.~Fraenkel\wei$^\dag$,
\footnotetext{$^\dag$Deceased}
C.~Garabatos\gsi,
P.~Gl\"{a}ssel\hei,
G.~Hering\gsi,
J.~Holeczek\gsi,
M.~Kalisky\mue,
S.~Kniege\fra,
V.~Kushpil\rez,
A.~Maas\gsi,
A.~Mar\'{\i}n\gsi,
J.~Milo\v{s}evi\'{c}\hei,
D.~Mi\'{s}kowiec\gsi,
R.~Ortega\hei,
Y.~Panebrattsev\dub,
O.~Petchenova\dub,
V.~Petr\'{a}\v{c}ek\hei,
M.~P\l{}osko\'{n}\fra,
S.~Radomski\hei,
J.~Rak\gsi,
I.~Ravinovich\wei ,
P.~Rehak\bnl,
H.~Sako\gsi,
W.~Schmitz\hei,
S.~Schuchmann\fra,
J.~Schukraft\cer,
S.~Sedykh\gsi,
S.~Shimansky\dub,
R.~Soualah\hei,
J.~Stachel\hei,
M.~\v{S}umbera\rez ,
H.~Tilsner\hei,
I.~Tserruya\wei,
G.~Tsiledakis\gsi,
J.\thinspace P.~Wessels\mue ,
T.~Wienold\hei,
J.\thinspace P.~Wurm\mpi,
S.~Yurevich\gsi,
V.~Yurevich\dub}

\affiliation{
\rez NPI ASCR, \v{R}e\v{z}, Czech Republic\\
\dub JINR Dubna, Russia\\
\gsi GSI Darmstadt, Germany\\
\fra Frankfurt University, Germany\\
\mpi MPI, Heidelberg, Germany\\
\hei Heidelberg University, Germany\\
\wei Weizmann Institute, Rehovot, Israel\\
\mue M\"{u}nster University, Germany\\
\sun SUNY Stony Brook, U.S.A.\\
\bnl BNL, Upton, U.S.A.\\
\cer CERN, Geneva, Switzerland\\
}

\date{\today}

\begin{abstract}
We present results of a two-pion correlation analysis performed with the 
Au+Pb collision data collected by the upgraded CERES experiment in the 
fall of 2000. 
The analysis was done in bins of the reaction centrality and the pion 
azimuthal emission angle with respect to the reaction plane. 
The pion source, deduced from the data, is slightly elongated in the 
direction perpendicular to the reaction plane, similarly as was observed 
at the AGS and at RHIC. 
\end{abstract}

\maketitle

%%%%%%%%%%%%%%%%%%%%%%%%%%%%%%%%%%%%%%%%%%%%%%%%%%%%%%%%%%%%%%%%%%%%%%%%%%%%%%%

\section{Introduction}\label{section:introduction}

Two-particle correlations provide unique access to the spatial 
extension of the source of particles emitted in the course 
of heavy ion collisions at relativistic energy (for a recent 
review see \cite{mike-overview}). 
The relation between the experimental correlation function, defined as 
the relative momentum distribution of pairs, normalized to the analogous 
distribution obtained via event mixing, and the size of the fireball is 
especially simple in case of identical pions where the main source of 
correlations is the Bose-Einstein statistics. 
In fact, in the pure boson case the correlation function is 
\begin{equation}\label{eq:correlation}
C(\mathbf{q}, \mathbf{P}) = 1 + \frac{|\int d^{4}x\, S(x, P)\,
\exp(iq\cdot x)|^2}{|\int d^{4}x \, S(x, P)|^2}\, 
\end{equation}
with the source function $S(x,P)$ describing the single particle density in 
8-dimensional position-momentum space at freeze-out. 
The correlation function $C$ depends on the momentum difference {\bf q} = 
{\bf p$_2$} - {\bf p$_1$} and the mean pion momentum 
{\bf k} = ({\bf p$_1$}+{\bf p$_2$})/2. 
The width of the peak at {\bf q} = 0 is inversely proportional to the source 
radius. A particularly exciting prospect is to look for a source asymmetry 
possibly reflecting the initial asymmetry of the fireball created in 
collisions with finite impact parameter. 
Indeed, significant dependence of the source radii on the azimuthal 
emission angle with respect to the reaction plane, defined by the beam 
axis and the impact parameter vector, was observed in Au+Au collisions at 
2-6 GeV \cite{ags} and at $\sqrt{s}$=200 GeV \cite{rhic200}. 
In this paper we present results of the first analysis of the azimuthal 
dependence of pion source radii at SPS energies. 

%%%%%%%%%%%%%%%%%%%%%%%%%%%%%%%%%%%%%%%%%%%%%%%%%%%%%%%%%%%%%%%%%%%%%%%%%%%%%%%
\section{Experiment}\label{section:experiment}
The CERES/NA45 experiment at the CERN SPS is described in detail in 
\cite{ceres}. 
The upgrade by a radial Time Projection Chamber (TPC) in 1997-1998, 
in addition to improving the dilepton mass resolution, 
enhanced the experiment's capability of studying hadronic observables. 
The cylindrical symmetry of the experiment was preserved during the upgrade, 
making the setup ideally suited to address azimuthal anisotropies. 
About 30 million Pb+Au collision events at 158\agev\ were collected in 
the fall of 2000, 
most of them with centrality within the top 7\% of the geometrical cross 
section \sg. 
Small samples of $\sigma$/\sg=20\% and minimum bias collisions, as well as 
a short run at 80\agev, were recorded in addition. 

%%%%%%%%%%%%%%%%%%%%%%%%%%%%%%%%%%%%%%%%%%%%%%%%%%%%%%%%%%%%%%%%%%%%%%%%%%%%%%%
\section{Data analysis}\label{section:analysis}
The results presented here are based on a correlation analysis of the 
high-statistics Pb+Au data set from the year 2000 \cite{darek}. 
The results of an azimuthal-angle averaged analysis in \cite{darek} are 
consistent with the previously published CERES data \cite{heinz-pap} and with 
a recent analysis by the NA49 Collaboration \cite{kniege}.  
The main steps of the azimuthal-angle dependent analysis are described below.

\subsection{Event selection}

The collision centrality was determined via the charged particle multiplicity 
around midrapidity $y_{\rm beam}$/2=2.91. 
Two variables, the amplitude of the Multiplicity Counter (MC) (single 
scintillator covering 2.3$<\eta<$3.5) and the track multiplicity in the 
TPC (2.1$<\eta<$2.8), were alternatively used as the centrality measure. 
Knowing the data acquisition dead time factor and the target thickness, 
and assuming that all beam particles were hitting the target, the event 
counts were translated to the cross section for collisions with a given 
multiplicity. 
The centrality variable used in this paper is defined as the integrated 
cross section divided by the geometrical cross section \sg=6.94~barn. 

The fireball created in a collision with finite impact parameter is elongated 
in the direction perpendicular to the reaction plane. In the course of 
expansion, with the pressure gradient larger in-plane than out-of-plane, 
the initial asymmetry should get reduced or even reversed. 
Experimentally, the source eccentricity at the time of decoupling can 
be determined from an analysis of two-pion correlations as function 
of $\Phi^*=\Phi_{\rm pair}-\Psi_{\rm RP}$, 
the pair emission angle with respect to the reaction plane. 

The azimuthal angle of the reaction plane $\Phi_{\rm RP}$ was estimated by the 
preferred direction of the particle emission (elliptic flow). 
For this, in each event the flow vector ${\bf Q}_2$ was constructed out of the 
measured particles, characterized by transverse momenta $p_t$ and azimuthal 
emission angles~$\phi$ \cite{danielewicz,e877}: 
\begin{eqnarray}
Q^X_2 &=& \frac{1}{N} \sum_{i=1}^N p_{t,i} \, \cos(2\phi_i)\\
Q^Y_2 &=& \frac{1}{N} \sum_{i=1}^N p_{t,i} \, \sin(2\phi_i) \, .
\end{eqnarray}
The $Q^X_2$ and $Q^Y_2$ components were calibrated (shifted and scaled) such 
that the peak in the $(Q^X_2,Q^Y_2)$ distribution was centered at (0,0), 
and its widths in the directions of $X$ and $Y$ were equal. 
The reaction plane angle was calculated (modulo $\pi$) from the calibrated 
${\bf Q}_2$  via 
\begin{equation}
\Phi_{\rm RP} = \frac{1}{2} \arctan \frac{Q^Y_2}{Q^X_2} .
\end{equation}
The resolution of the so determined reaction plane angle, estimated via the 
subevent method \cite{danielewicz, poskanzer}, was 30-34\degrees. 

The event mixing, needed to obtain the denominator of the correlation 
functions, 
was performed in bins of centrality (2\% of \sg), event plane (15\degrees), 
and vertex (same target disk).  
This ensures that the shape of the background is identical to that of the 
signal in all respects except for the femtoscopic correlations. 

\subsection{Track selection}
Only TPC tracks with at least 12 (out of maximally 20) hits and 
a reasonably good $\chi^2$, falling into the fiducial acceptance 
$0.125<\theta<0.240$, were used in the analysis. 
A momentum dependent d$E$/d$x$ cut was applied to reduce the contamination of 
the pion sample. 
Pions from $K^{0}$ and $\Lambda^{0}$ decays were suppressed by a 2.5 $\sigma$ 
matching cut between the silicon vertex detectors and the TPC. 

\subsection{Pair selection}
The two-track resolution cuts applied to the true pairs and to the pairs 
from event mixing were different for the two possible track pair topologies: 
%(Fig.~\ref{fig:sailor-cowboy}). 
in the case of the magnetic field bringing the tracks apart from each other 
(sailor) the required two-track separation was $\Delta \phi>$38-45~mrad, 
depending on the transverse momentum; 
for the opposite (cowboy) case, $\Delta \phi>$90-140~mrad was used. 
The polar angle cut was the same for both topologies 
($\Delta \theta\!\!>$8-9~mrad). 
It should be noted that the required two-track cuts depended somewhat on the 
quality cuts applied to single tracks: the higher the number of hits required 
for single tracks, the more pairs were lost because of the finite two-track 
resolution. 

\subsection{Correlation functions}
The two-pion analysis was performed in the longitudinally co-moving frame 
(LCMS) defined by the vanishing $z$-component of the pair momentum. The 
momentum difference in this frame, {\bf q} = {\bf p}$_2$-{\bf p}$_1$, was 
decomposed into the ``out'', ``side'', and ``long'' components following the 
Bertsch-Pratt convention, with $q_{\rm out}$ pointing along the pair 
transverse momentum and $q_{\rm long}$ along the beam \cite{bertsch}. 
The particles were numbered such that \qside\ was always positive. 

The \pim\pim\ and \pip\pip\ correlation functions were fitted by 
\begin{eqnarray}
C_2\left({\bf q}\right) &=& N \cdot\left\{\left(1-\lambda\right) + 
\lambda \cdot F_{c}\left({\bf q}\right)(1+G({\bf q}))\right\} , \nonumber \\
G({\bf q}) & = & 
\exp(-\displaystyle\sum_{i,j} R^{2}_{ij} 
q_{i}q_{j})
\label{eq:fitfunction}
\end{eqnarray}
with the indices $i$,$j$ = \{out, side, long\}. 
The normalization factor $N$ was needed because the number of pairs from 
event mixing was four times higher than of signal pairs. 
The correlation strength $\lambda<1$ reflects 
the contribution of pions from long-lived resonances, 
the contamination of the pion sample by other particle species, 
and the finite ${\bf q}$-resolution. 
The resulting source radii $\sqrt{R^{2}_{ij}}$ describe the size of the source 
emitting pions of a given momentum \cite{mike-overview}. 
The $F_{c}\left(q_{inv}\right)$ factor, $q_{inv}=\sqrt{-(p_2^\mu-p_1^\mu)^2}$, 
accounts for the mutual Coulomb interaction between the pions
and was calculated by averaging the nonrelativistic Coulomb wave function 
squared over a realistic source size. 
The Coulomb factor was attenuated by $\lambda$ similarly as the rest of the 
correlation function peak; the importance of this approach was demonstrated 
in \cite{heinz-pap}. 
The fits were performed by the minimum negative log-likelihood method, 
assuming that the number of true pairs in a given bin is distributed 
around the expected mean value (equal to the number of mixed pairs times 
the fit function) according to the Poisson statistics. 
%and were done separately for each pair (\pt,y) bin. 
The source radii obtained from the fit were corrected for the finite 
momentum resolution 
\begin{equation}
\frac{\Delta p}{p} = 2\% \oplus 1\% \cdot p/{\rm (GeV\!/c)} .
\end{equation}
The correction was determined by Monte Carlo and is rather insignificant 
for \Rside\ and \Rlong ; for \Rout\ it gets as large as $\approx$ +20\% for 
the highest pion momenta. 

\subsection{Azimuthal dependence of pion source radii}

Correlation functions were generated separately for pion pairs with 
different azimuthal angles with respect to the reaction plane 
$\Phi^*=\Phi_{\rm pair}-\Psi_{\rm RP}$. 
For this, the pions were sorted into eight bins covering $(-\pi/2,\pi/2)$. 
The eight correlation functions were fitted \footnote{
To avoid problems caused by the non-gaussian shape of inclusive correlation 
functions the fit was actually performed in bins of pair \pt; the $R_{i,2}$'s 
discussed in this paper are weighted averages of the $R_{i,2}$'s 
obtained for different \pt's.} 
and the resulting squared source radii were parametrized by \cite{heinz-lisa} 
\begin{eqnarray}
\rout^2  &=& {\rout^2}_{,0}  + 2 {\rout^2}_{,2} \cos(2 \Phi^*) \nonumber \\
\rside^2 &=& {\rside^2}_{,0} + 2 {\rside^2}_{,2} \cos(2 \Phi^*) \nonumber \\
\rlong^2 &=& {\rlong^2}_{,0} + 2 {\rlong^2}_{,2} \cos(2 \Phi^*) \nonumber \\
\routside^2  &=& 2 {\routside^2}_{,2}  \sin(2 \Phi^*) \nonumber \\
\routlong^2  &=& {\routlong^2}_{,0} + 2 {\routlong^2}_{,1}  \cos(  \Phi^*) 
                  \nonumber \\
\rsidelong^2 &=& 2 {\rsidelong^2}_{,1} \sin(  \Phi^*) .
\label{eq:rfourier}
\end{eqnarray}
While the $R_{\rm out,0}^2$, $R_{\rm side,0}^2$, and $R_{\rm long,0}^2$ 
obtained represent the phi-averaged squared pion HBT radii, 
the second Fourier coefficients $R_{i,2}^2$ describe the eccentricity of 
the observed pion source. 
Since the reaction plane is known modulo $\pi$ odd Fourier components 
should vanish. 
If the pion source were to reflect the initial collision geometry (almond
shape out-of-plane) a positive ${\rside^2}_{,2}$ and ${\routside^2}_{,2}$ 
and a negative ${\rout^2}$ should be expected. 
For symmetry reasons all anisotropies should disappear in the limit of central 
collisions. 

The second Fourier coefficients $R_{i,2}^2$ have been corrected for the 
reaction plane resolution by dividing them by the mean cosine of twice 
the difference between the reconstructed and the true reaction plane angles 
$\langle \cos[2 (\Psi_{\rm RP}^{\rm rec}-\Psi_{\rm RP})] \rangle$, 
estimated via the measured mean cosine of the difference between 
two subevents: 
\begin{equation}
R_{i,2}^2 \rightarrow \frac{R_{i,2}^2}{\sqrt{2 \langle \cos [2 (\Psi_b-\Psi_a)]\rangle}} , 
\label{eq:rpres}
\end{equation}
similarly as it is done for flow measurements \cite{poskanzer}. 
The correction factor was between 4.8 and 2.4 (centralities 
of 0-2.5\% and 25-70\%, respectively). 
The appropriateness of the flow correction for two-pion correlation 
radii has been verified with a Monte Carlo simulation of pion pairs 
emitted from an elliptical source;  
the observed reduction of the second Fourier coefficients of squared radii 
was, within the statistical errors of the simulation (about 10\%), 
in agreement with Eq.~(\ref{eq:rpres}). 

\subsection{Systematic errors}
The fact that separate analyses of positive and negative pions give 
consistent results indicates that the limited particle identification is 
not causing any bias. 
The systematic error related to the reaction plane resolution correction, 
based on the comparison between Eq.~(\ref{eq:rpres}) and a  
numerical simulation, is estimated to be not larger than 10\%. 
Autocorrelations, i.e. using the same particles to determine the event plane 
and the HBT radii, might be another source of systematic errors; 
unfortunately, an attempt to exclude the two particles from the event plane 
construction led to a strong distortion of the correlation functions as it was 
not clear how to do the exclusion consistently for the true and mixed pairs. 
An independent way to estimate the overall systematic errors is by inspecting 
(see Fig.~\ref{fig:r2} in the next section) the coefficients 
\Routlong$_{,1}^2$ and \Rsidelong$_{,1}^2$ which should be zero because 
the reaction plane is known only modulo $\pi$. 
Averaged over centralities, they are 0.40 (16) and 0.22 (6) fm$^2$, 
respectively. 
Attributing part (one sigma) of the deviation to statistical fluctuation 
we are left with a discrepancy of about 0.3 fm$^2$ which we thus assume to 
be the systematic uncertainty of the results presented. 

%%%%%%%%%%%%%%%%%%%%%%%%%%%%%%%%%%%%%%%%%%%%%%%%%%%%%%%%%%%%%%%%%%%%%%%%%%%%%%%
\section{Results and discussion}

\begin{figure}[b]
\hspace*{-6mm}\includegraphics{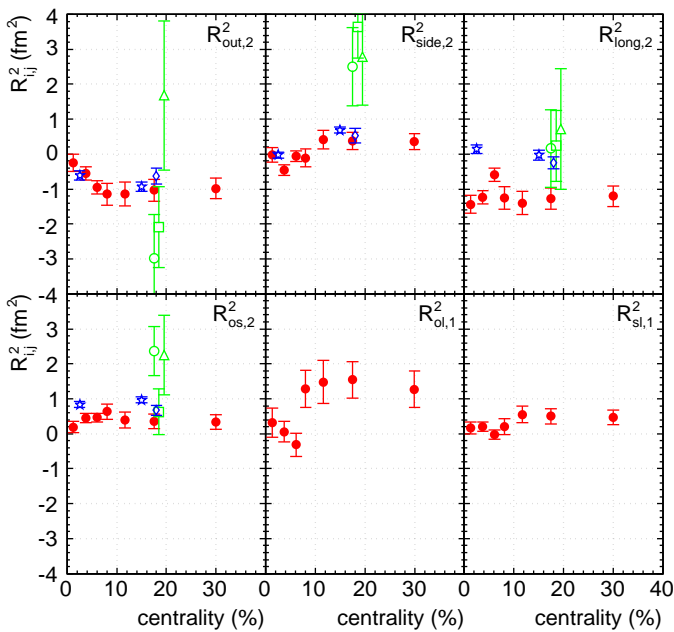}
\caption{
Azimuthal pion source eccentricity, 
represented by the second Fourier coefficient of squared radii 
$R_i^2(\Phi^*)$, measured in Pb+Au collisions as a function of centrality. 
Positive and negative pion pairs have been combined. 
The mean pion transverse momentum is 0.23~GeV/c. 
For comparison, 
analogous measurements at the AGS (open green symbols) and RHIC 
(blue diamonds and stars for 130 and 200 GeV, respectively) are shown. 
The last two panels show the first Fourier coefficients ${\routlong^2}_{,1}$ 
and ${\rsidelong^2}_{,1}$ which should be zero by construction. 
The estimated systematic error is 0.3 fm$^2$. 
}
\label{fig:r2}
\end{figure}

The extracted Fourier coefficients are shown in Table~\ref{table:centrality} 
and Fig.~\ref{fig:r2} \footnote{
We refrain in Fig.~\ref{fig:r2} from normalizing the coefficients 
to the corresponding mean source radii. 
In fact, since the second Fourier coefficients measured at RHIC 
\cite{rhic200} seem to be, within the measurement errors, independent 
of transverse momentum while the mean radii vary by about 40\%, 
normalization to mean radii could wash out the weak oscillation we 
are after.}.
\begin{table*}[t!]
\caption{\label{table:centrality}
Fourier coefficients from a fit of Eq.~(\ref{eq:rfourier}) to 
the pion source radii squared.
The values are in fm$^2$. }
\begin{ruledtabular}
\begin{tabular}{cccccccccc}
centrality & mean cent 
& ${\rout^2}_{,0}$ & ${\rside^2}_{,0}$ & ${\rlong^2}_{,0}$ & ${\routlong^2}_{,0}$ 
& ${\rout^2}_{,2}$ & ${\rside^2}_{,2}$ & ${\rlong^2}_{,2}$ & ${\routside^2}_{,2}$ \\
% pt-inclusive fit
%0-2.5\%  &   1.3 \%  &   29.34(10) &  24.33(08) &  33.10(11) &  -6.56(21)   &  -0.44(32) &  -1.18(25) &  -2.15(35) &   0.14(23)   \\
%2.5-5\%  &   3.7 \%  &   28.01(08) &  23.29(06) &  31.91(09) &  -5.72(18)   &  -0.84(24) &  -0.55(18) &  -2.01(27) &   0.73(15)   \\
%5-7.5\%  &   6.1 \%  &   26.62(08) &  22.04(06) &  30.61(10) &  -5.22(15)   &  -1.18(20) &  -0.68(15) &  -1.61(24) &   0.84(13)   \\
%7.5-10\% &   8.1 \%  &   25.39(17) &  21.00(13) &  29.22(21) &  -5.20(31)   &  -1.36(37) &   0.28(28) &  -1.68(45) &   0.85(28)   \\
%10-15\%  &  11.6 \%  &   23.83(19) &  19.87(14) &  27.31(23) &  -4.33(37)   &  -0.88(37) &   0.07(28) &  -0.68(45) &   0.80(26)   \\
%15-25\%  &  17.5 \%  &   21.35(23) &  17.62(17) &  24.64(28) &  -3.67(48)   &  -1.43(41) &  -0.63(30) &  -0.53(48) &   1.09(27)   \\
%25-70\%  &  30.0 \%  &   14.28(49) &  12.29(35) &  17.42(61) &  -4.23(116)  &  -0.55(86) &   0.51(58) &   1.19(102) &   1.15(51)  
% fit in bins of pt, averaged
0-2.5 \%  &   1.3 \%  &   29.34(10) &  24.33(08) &  33.10(11) &  -6.56(21)   &  -0.25(25) &  -0.03(21) &  -1.43(26) &   0.19(17)  \\
2.5-5 \%  &   3.7 \%  &   28.01(08) &  23.29(06) &  31.91(09) &  -5.72(18)   &  -0.56(19) &  -0.45(15) &  -1.23(19) &   0.45(13)  \\
5-7.5 \%  &   6.1 \%  &   26.62(08) &  22.04(06) &  30.61(10) &  -5.22(15)   &  -0.96(19) &  -0.06(15) &  -0.59(19) &   0.46(12)  \\
7.5-10 \% &   8.1 \%  &   25.39(17) &  21.00(13) &  29.22(21) &  -5.20(31)   &  -1.15(32) &  -0.11(26) &  -1.25(32) &   0.63(21)  \\
10-15 \%  &  11.6 \%  &   23.83(19) &  19.87(14) &  27.31(23) &  -4.33(37)   &  -1.14(35) &   0.42(27) &  -1.40(34) &   0.39(23)  \\
15-25 \%  &  17.5 \%  &   21.35(23) &  17.62(17) &  24.64(28) &  -3.67(48)   &  -1.03(31) &   0.38(24) &  -1.27(30) &   0.35(21)  \\
25-70 \%  &  30.0 \%  &   14.28(49) &  12.29(35) &  17.42(61) &  -4.23(116)  &  -0.98(30) &   0.36(23) &  -1.20(29) &   0.34(20)  
\end{tabular}
\end{ruledtabular}
\end{table*}
The anisotropies in the out and side directions indicate a pion source 
elongated out-of-plane.  
For comparison, the AGS \cite{ags} and RHIC \cite{rhic130,rhic200} 
values, obtained by performing the fits using Eq.~(\ref{eq:rfourier}) on their 
published radii (the AGS results were subsequently corrected for their 
reaction plane resolution),  
are represented in Fig.~\ref{fig:r2} by open symbols and stars.
It appears that the anisotropies in the out and side directions at SPS 
energy are rather similar to those observed at RHIC. 
The geometrical pion source eccentricity can be quantified via 
$\varepsilon \simeq 2 {\rside^2}_{,2}/{\rside^2}_{,0}$ 
and has, for a centrality of 15-20\%, a value of 0.043(27), 
significantly less than the initial fireball 
eccentricity $\varepsilon_{\rm initial} \approx 0.20$ \cite{rhic200}. 
The fact that the magnitude of the \Rside\ anisotropy seems to be 
weaker than that of \Rout\ indicates that not only the source geometry 
but also azimuthal dependence of the emission time might play a role. 
On the other hand, 
${\rout^2}_{,2} - {\rside^2}_{,2}$+2${\routside^2}_{,2}$, 
averaged over centralities, is 0.06(18)~fm$^2$, consistent with zero. 
This implies that the sum rule from \cite{heinz-lisa}, 
which is supposed to be valid for systems with emission time 
independent on the azimuthal angle, 
works rather well. 

The \Rlong\ anisotropy is negative and roughly independent of 
centrality. 
Series of checks were performed to make sure that this is not an 
artefact of the analysis. This result might indicate that \Rlong\ 
is sensitive to fluctuations of the azimuthal particle density. 
Hydrodynamic calculation of central Pb+Au collisions at this energy 
yields a similar amount of anisotropy in $R_{\rm long,2}^2$ 
which, however, is centrality dependent \cite{pasi}. 
This, and the fact that the hydrodynamic calculation overpredicts the 
overall magnitude of \Rlong, indicates that the knowledge about the 
mechanism leading to oscillations of \Rlong\ is still incomplete. 

\begin{figure}[t]
\hspace*{-5mm}\includegraphics[clip=]{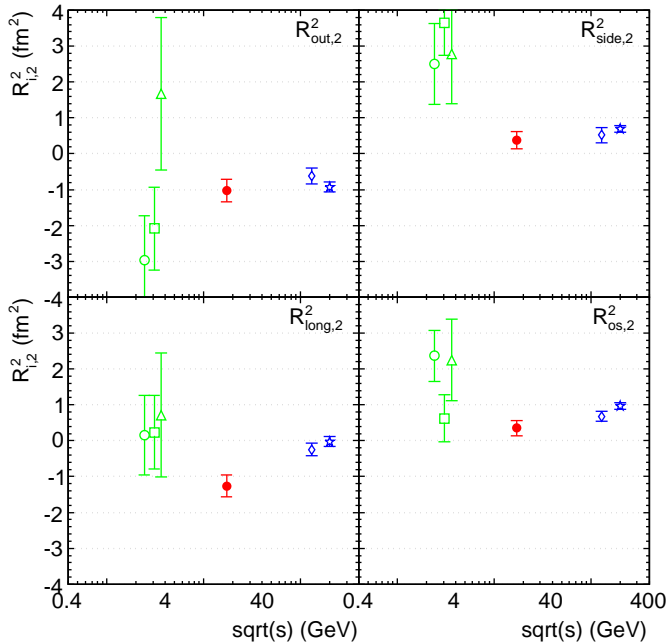}
\caption{
Collision energy dependence of the pion source anisotropy in Au+Au and 
Pb+Au collisions at the 15-20\% centrality. 
The meaning of the symbols is the same as in Fig.~\ref{fig:r2}. 
The SPS results are rather similar to those obtained at RHIC, except 
for \Rlong which is significantly off-zero. 
The estimated systematic error is 0.3 fm$^2$. 
\label{fig:vssqrts}
}
\end{figure}

The source radius anisotropies for centrality 15-20\% are shown  
as a function of the collision energy in Fig.~\ref{fig:vssqrts}. 
The SPS result fits rather well into the beam energy systematics. 
The apparent (albeit not statistically significant) fast 
change of ${\rside^2}_{,2}$ between AGS and SPS and, even more,  
the negative ${\rlong^2}_{,2}$ developing when going down in energy 
from RHIC to SPS, make the perspective of a systematic study in the 
course of the low-energy scan at RHIC especially attractive. 
With the statistical errors of the present AGS data even significant 
structures in the energy dependence of the pion source anisotropy 
cannot be excluded. 

%%%%%%%%%%%%%%%%%%%%%%%%%%%%%%%%%%%%%%%%%%%%%%%%%%%%%%%%%%%%%%%%%%%%%%%%%%%%%%%
\section{Summary}
We have analyzed the azimuthal angle dependence of the pion HBT radii in Pb+Au 
collisions at the top SPS energy. 
The source anisotropy in the out and side directions has the same sign and 
similar magnitude as the one measured at the AGS and 
at RHIC, and indicates a pion source elongated out-of-plane. 
The side anisotropy is somewhat smaller than the other which suggests that 
finite emission times may play a role. 
The source anisotropy in the long-direction is negative for all centralities 
indicating that \Rlong\ might be sensitive to particle density fluctuations. 

The CERES collaboration acknowledges the good performance of the CERN
PS and SPS accelerators as well as the support from the EST
division. We would like to thank R.~Campagnolo, L.~Musa, A.~Przybyla, 
W.~Seipp and B.~Windelband for their contribution during construction 
and commissioning of the TPC and during data taking.
We are grateful for excellent support by the CERN IT division for the 
central data recording and data processing. 
This work was supported by GSI, Darmstadt, the German BMBF, 
the German VH-VI 146, the US DoE, the Israeli Science Foundation, 
the Check Science Foundation contract No. 202/03/0879 and the 
MINERVA Foundation.

%%%%%%%%%%%%%%%%%%%%%%%%%%%%%%%%%%%%%%%%%%%%%%%%%%%%%%%%%%%%%%%%%%%%%%%%%%%%%%%

\end{document}